\documentclass[11pt]{article}
\usepackage[utf8]{inputenc}
\usepackage[english]{babel}
\usepackage[margin=2.3cm]{geometry} %
\usepackage{graphicx} %
\usepackage{amsthm, amsmath, amssymb} 
\usepackage{tcolorbox} 
\usepackage[numbers]{natbib} %
\usepackage{lineno}	
\usepackage{appendix} %
\usepackage{textcomp}
\usepackage{gensymb}
\usepackage{comment}
\usepackage{subcaption}
\usepackage{setspace}
\usepackage{siunitx}

\usepackage{authblk} 
 \usepackage{setspace} %
\usepackage{pifont}

\usepackage{hyperref}
\usepackage{cleveref} %
\newcommand{\papertitle}{Breaking better: How imperfections increase fracture resistance in architected lattices}

\newcommand{\ie}{{\textit{i.e.}}}
\newcommand{\eg}{{\textit{e.g.}}}

 \graphicspath{ {./} }
\begin{document}

\title{\papertitle}
\author[1]{Alessandra Lingua}
\author[1]{Antoine Sanner} %
\author[2]{François Hild} %

\author[1]{David S. Kammer\thanks{Corresponding Author: dkammer@ethz.ch}}

\affil[1]{Institute for Building Materials, ETH Zurich, Switzerland}
\affil[2]{Université Paris-Saclay, CentraleSupélec, ENS Paris-Saclay, CNRS\\ LMPS--Laboratoire de Mécanique Paris-Saclay, Gif-sur-Yvette, France}
 
\maketitle
\section*{Abstract}
\begin{comment}
Architected materials offer unique opportunities to tailor fracture properties through local structural modifications.
However, identifying the resulting failure mechanisms (\eg~crack path deviation, pinning, crack trapping) is complex, as they are highly localized and influenced by multiple factors including geometry, material heterogeneity, and possible defects. %
By leveraging the introduction of vacancies as a potential toughening element, which represent well-controlled and localized imperfections, we assess whether the work to failure of irregular, brittle lattices increases.
%
Using a combination of macroscopic mechanical testing and digital image correlation (DIC), we investigate which toughening mechanisms are triggered by the interaction between imperfections and a propagating crack.  We observe that these designed imperfections do not affect the failure initiation site or the peak tensile load. However, they activate crack propagation mechanisms that significantly increase the work to failure. DIC tracking and direct image inspection reveal that these toughening mechanisms include crack path tortuosity and crack bridging, depending on the location of the removed lattice beams. These results demonstrate that small, well-characterized imperfections, when properly mastered, can activate crack propagation mechanisms that enhance the failure resistance and broaden the design space of architected materials beyond regular, periodic structures.
\end{comment}
%
Fracture behavior in architected materials can be influenced by heterogeneities, yet the mechanisms by which imperfections affect crack propagation remain poorly understood. In this study, we introduce well-controlled, localized defects in the form of isolated missing struts to evaluate their impact on crack growth in brittle lattice specimens. Using mechanical testing combined with digital image correlation (DIC), we track crack propagation and identify failure processes at the scale of individual lattice cells. While the imperfections do not alter the location of crack initiation or the peak load, they consistently lead to an increase in work to failure when the crack path is tortuous or crack bridging occurs. These findings demonstrate how small, targeted modifications to an otherwise regular lattice can significantly influence fracture resistance in brittle architected materials.

\vspace{5mm}\noindent \textbf{Keywords:}
Architected materials; Additive manufacturing; Fracture characterization; Digital image correlation (DIC); Imperfections
\section*{Nomenclature} \label{sec:nomenclature}
DIC~ Digital image correlation\\
CT~ Compact tension\\
$w$~ Specimen width\\
$b$~ Specimen thickness\\
$a$~ Pre-crack length\\
$\sigma_u$~ Ultimate tensile strength\\
ROI~ Region of interest\\
$P_{max}$~ Peak force\\
$K_{IC}$~ Mode I stress intensity factor\\
$q$~ Auxiliary function for the calculation of $K_{IC}$\\
$W$~ Work to failure \\
$W_{el}$~ Elastic contribution of the work to failure\\
$P$~ Force\\
$d$~ Grips displacement\\
$d_u$~ Ultimate displacement \\
$\rho_u$~ Correlation gray-level residuals\\
$f(x)$~ Reference image\\
$g(x)$~ Deformed image\\
$U(x)$~ DIC displacement\\
$\varepsilon_1$~ DIC principal strain\\
$\Delta y$~ Crack path tortuosity\\
$h$~ Cell height\\
$x_c$~ x coordinate of the element centroids\\

\section{Introduction} \label{sec:introduction}

Rapid advances in additive manufacturing have enabled the fabrication of architected materials with exceptional mechanical properties, including auxetic behavior~\cite{SHAN2015}, local strengthening~\cite{tancogne_2018}, reprogrammable stiffness~\cite{zheng_2014}, and enhanced toughness~\cite{bertoldi_2017, liu_2020}. Among these materials, periodic lattices are particularly promising for applications that demand low density and tailorable mechanical properties, such as in aerospace or biomedical engineering~\cite{zhang_2015, Ongaro_2018}. Their discrete architecture offers a unique opportunity to control fracture behavior through topological design, potentially enabling improved damage tolerance~\cite{LIU2025}. Yet, despite this potential, fracture resistance remains one of the least understood mechanical properties of architected materials. 

While it is well established that the global failure of lattices results from the progressive damage of discrete elements, typically struts or nodes~\cite{Gibson_Ashby_1997, fleck_2010, quintana_2009}, the link between individual failure events and the formation and propagation of a macroscopic crack remains difficult to establish, both experimentally and numerically~\cite{montemayor_2016, JOST2021}. 
Although numerical studies have characterized the toughness of lattices~\cite{Yavas_2024,seiler_2019, CHOUKIR2025}, they do not fully capture the relationship between local behavior and global fracture response. On the experimental side, this gap is even more pronounced. The discrete structure of lattices complicates the tracking of crack propagation~\cite{RADI2023}, limiting our ability to identify the local mechanisms responsible for increased toughness in architected materials.

Identifying the dominant failure mechanisms is essential for designing engineered lattices with enhanced toughness~\cite{Hsieh_2020}. To this end, several approaches have been explored. These include, for instance, semi-regular tessellations with two vertex configurations to tailor the global fracture response~\cite{omidi_2023}. A further promising approach, inspired by composite materials, consists of combining two distinct phases for increased fracture toughness~\cite{fox_2025,TANKASALA2020}. Another strategy introduces uniformly distributed disorder by randomizing node positions~\cite{BHUWAL2023} or connectivity~\cite{karapiper_2023, romijn_2007, zaiser_2023}, often drawing inspiration from the imperfect and hierarchical structures of biomaterials~\citep{barthelat_2007, Gao_24, manno_2019, Bhuwal_2023, vangelatos_2023}. These disordered lattices activate mechanisms such as crack deflection, diffused damage~\cite{fulco2025}, and crack bridging~\cite{TANKASALA2020}, which can contribute to increased toughness. However, in all these studies, many defects are present simultaneously, which hinders the identification of the main mechanisms responsible for the toughening~\cite{CHOUKIR2023}. 
A promising strategy to uncover the local failure mechanisms in imperfect lattices is to introduce \emph{localized} disorder, such as material heterogeneities, bond strength variations, geometric irregularities, or vacancies. Because these imperfections are spatially confined, they allow for clearer identification of the mechanisms they activate, in contrast to uniformly distributed disorder, which alters the structure globally and makes causal links harder to establish~\cite{karapiper_2023, fulco2025}. In the context of 3D printing, naturally occurring imperfections are abundant and randomly distributed throughout the material~\cite{BENEDETTI2021, li_2021, Maurizi_2022}. Introducing a controlled imperfection, such as the removal of a single strut (\ie{}, introducing a vacancy), offers a distinct advantage because it creates a localized and reproducible defect that can be systematically studied against a regular baseline. Previous studies at the molecular and continuum scales suggest that such localized vacancies can increase fracture toughness~\cite{Brescakovic_2022, Urabe, zhao_2025, CURTIN19902051, deng}. However, these effects are not universal, and it remains unclear whether similar mechanisms arise in brittle architected materials. More importantly, the specific toughening mechanisms triggered by such imperfections have yet to be systematically identified.
Here, we experimentally investigate the fracture of brittle lattice materials and assess how localized failure mechanisms contribute to increased work to failure. To this end, we introduce isolated missing struts into lattice specimens and combine mechanical testing with full-field optical measurement. We employ digital image correlation (DIC) to track the crack tip, extract the crack path, and localize failure at the level of individual lattice cells. Our results reveal two toughening mechanisms that correlate with the increased work to failure observed in imperfect specimens: increased crack path tortuosity and crack bridging, where struts behind the crack tip remain temporarily intact.  
These findings clarify the link between local failure processes and global fracture response, enabling more systematic exploration of toughening strategies in architected materials.

\section{Fracture characterization} \label{sec:methods}

We characterize the fracture behavior of regular and imperfect lattices by testing 3D-printed compact tension specimens with and without vacancies under mode I loading.

\subsection{Design of specimens with controlled imperfections} \label{sec:specimen_design}

\begin{figure}[!h]
    \centering
    \includegraphics[width=.55\linewidth]{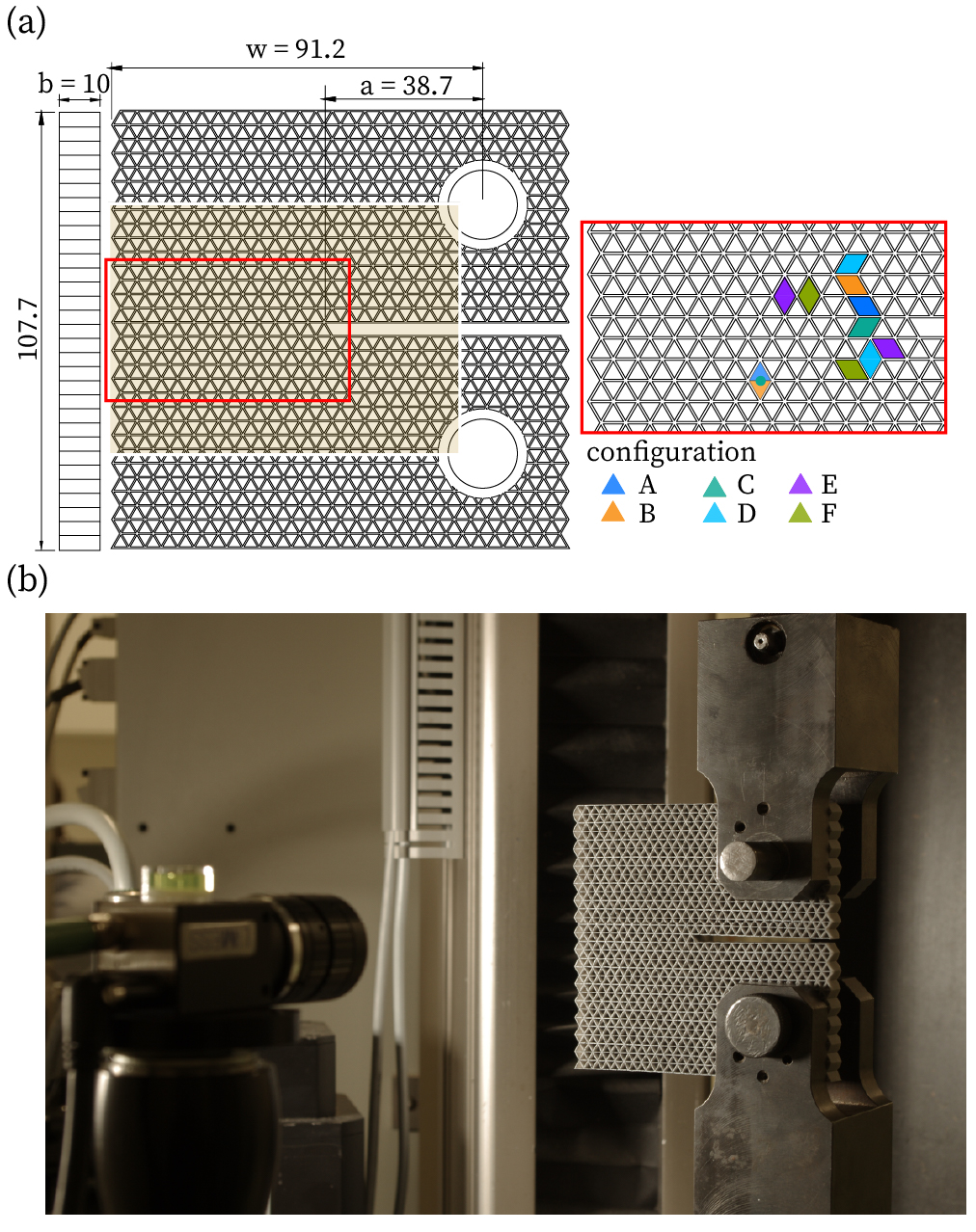}
    \caption{Design and setup of compact tension specimens for characterizing the fracture behavior of lattices. (a)~Geometry of the compact tension specimen and location of the missing struts in imperfect lattices. Configurations A to F differ in the position of missing struts, located at varying distances from the notch root. Dimensions are given in millimeters. (b)~Mechanical testing and imaging setup, which includes a LIMESS high-resolution camera for real-time imaging of the patterned specimen surfaces during fracture characterization.}
    \label{fig:1}
\end{figure}

We design compact tension (CT) specimens with a triangular lattice topology, adapted from the ASTM E-1820 standard~\cite{ASTM_E1820} (Fig.~\ref{fig:1}a). The unit cell width $l$ is set to $4~\mathrm{mm}$ and the strut width to 0.38~mm, resulting in a relative density of $\sim$0.33. This design represents a compromise between minimizing density, maximizing printing volume, and preserving specimen quality. In addition to regular lattice specimens, we fabricate specimens with controlled imperfections. Specifically, we introduce imperfections by removing two struts from the regular lattice design. The low number of controlled defects enables us to assess crack sensitivity to individual defects and evaluate their effects twice within a single test. In addition, it ensures that one missing strut is positioned above and the other below the notch root to maximize the chances of crack interaction with the imperfections.
We consider six imperfect configurations, labeled from A to F (Fig.~\ref{fig:1}a), each involving two missing struts in the vicinity of the notch but differing in their (absolute and relative) locations. 

We manufacture eight regular and fourteen imperfect specimens using a Form3 stereolithography machine with Gray V4 resin, both supplied by Formlabs. The printed material has an average elastic modulus of $2.7 \pm 0.5~\mathrm{GPa}$ and an ultimate tensile strength of $\overline{\sigma}_\mathrm{u} = 49.0\pm 3~\mathrm{MPa}$, as revealed by standard characterization of the resin. A layer thickness of $100~\micro\mathrm{m}$ provides an optimal balance between printing speed and part quality. Printing supports ease the specimen separation from the platform. The excess resin is then removed in the Form Wash with isopropanol, and the specimens are cured in the Form Cure for 30~min at 60$\degree$C with the printing support to minimize warping. To ensure consistency, we print all specimens in the same machine, with each print taking about 5~h. Following printing, the specimens were stored in the unplugged Form cure, which has a UV light shield, for up to 120~h (5 days) to prevent unmonitored light curing before testing.

\subsection{Mechanical testing and imaging} \label{sec:mechanical_testing}

The as-printed width ($w$), thickness ($b$), and pre-crack length ($a$) (Fig.~\ref{fig:1}a) are measured for each specimen to check their agreement with the nominal dimensions, accounting for potential warping and platform removal effects. The specimens are loaded in tension using a Zwick-Roell Z100 machine at a displacement-controlled rate of 2~mm/min, with a pre-load of 4~N. The machine records the load up to a 90\% peak load drop. As shown in Fig.~\ref{fig:1}b, a high-resolution LIMESS camera is placed in front of the specimen to track crack propagation over a region of interest (ROI) of approximately $85 \times 60$~mm$^2$. We lubricate the loading pins before testing to prevent stick-slip and induced rigid-body rotations of the specimen under load. The critical stress intensity factor is estimated according to the ASTM E-1820 standard~\cite{ASTM_E1820}, which provides an approximation of the lattice toughness accounting for slight variations in dimensions across samples. The mode I critical stress intensity factor ($K_\mathrm{IC}$) is then defined as
\begin{equation}
    K_{IC}=\frac{P_{\max}}{b\sqrt{w}} ~ q\left(\frac{a}{w}\right) ~,
    \label{eq:k_ex}
\end{equation} 
where $P_{\max}$ is the peak force recorded by the load cell of the testing machine, and $q$ is a function of the $a/w$ ratio specific to the specimen design~\cite{ASTM_E1820}. 

\section{Macroscopic properties} \label{sec:macroproperties}

We first assess the influence of imperfections on the onset of failure by analyzing the response of regular and imperfect specimens up to the peak load. We then quantify the effect of the designed vacancies on the work to (ultimate) failure to identify configurations of interest for in-depth analysis of the crack path in Sec.~\ref{sec:crack_path_characteristics}.

\subsection{Failure initiation} \label{sec:failure_init}

\begin{figure*}[!h]
    \centering
    \includegraphics[scale=.25]{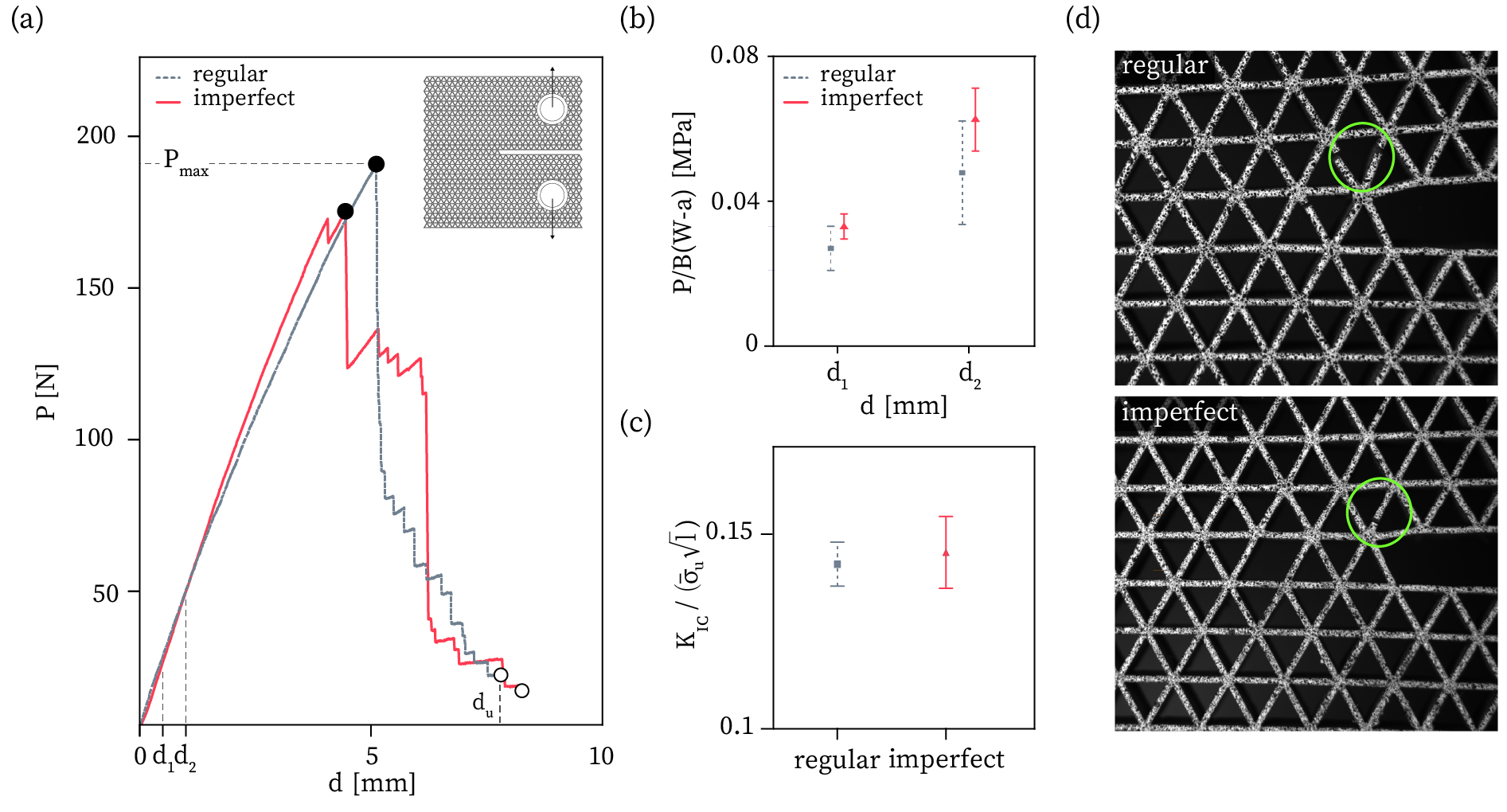}
    \caption{Failure initiation for regular and imperfect specimens is equivalent. (a)~Representative force-displacement curves for one regular specimen and one imperfect specimen. The peak forces $P_\mathrm{max}$ and ultimate displacement $d_\mathrm{u}$ are marked by solid and open black circles, respectively. (b)~Average normalized force (with 90\% confidence interval) in the elastic regime at macroscopic displacements of $d_1 = 0.5$ and $d_2=1$, as marked by vertical dashed lines in (a), based on eight regular and six imperfect specimens. (c)~Critical stress intensity factor $K_\mathrm{IC}$, computed using Eq.~(\ref{eq:k_ex}) and normalized by $\overline{\sigma}_\mathrm{u}$ (the ultimate stress of base material), and $\sqrt{l}$ (with unit cell width $l$). Error bars indicate 90\% confidence intervals. 
    (d)~Images of crack tip area for regular (top) and imperfect (bottom) specimens corresponding to those shown in (a), after the first struts fail. The first broken struts are highlighted with a green circle. The regular specimen shows two struts that failed simultaneously.
    }
    \label{fig:2}
\end{figure*}

We examine the macroscopic response of the regular and imperfect lattices under mode I loading. Both types of lattices exhibit similar force-displacement response, as shown for representative specimens in Fig.~\ref{fig:2}a. Initially, the response is nearly linear, extending up to the maximum force, $P_\mathrm{max}$. Beyond this point, the force drops abruptly, which is in line with the commonly observed brittleness of lattice materials~\cite{Fleck_2007, LUAN2022}. However, this force drop does not immediately lead to catastrophic failure of the specimen, and it is followed by step-wise load decay, which leads to significant additional displacement before ultimate failure.

The linear response of regular and imperfect lattices is quantitatively consistent, as reported in Fig.~\ref{fig:2}a. To enable a systematic comparison and account for variability due to warping, the force is normalized with the area $b(w-a)$ at two displacement levels $d$ in the quasi-linear domain. While some inconsistency in the elastic response of the lattices is observed, likely due to material variability and environmental conditions, the missing struts do not affect the elastic response in a statistically significant way (Fig.~\ref{fig:2}b).

We assess the resistance to failure initiation by computing the critical stress intensity factor using Eq.~(\ref{eq:k_ex}). The average values of $K_\mathrm{IC}$ for regular and imperfect specimens are statistically equivalent, as shown in Fig.~\ref{fig:2}c. The inspection of the images from the first failure event, associated with the force drop at $P_\mathrm{max}$, reveals that the same strut at the notch root breaks in both regular and imperfect specimens (Fig.~\ref{fig:2}d). This is true for almost all tested specimens (not shown). Thus, it is concluded that the missing struts in the six configurations considered herein do not alter the crack initiation mechanism, maintaining an equivalent critical stress intensity factor.

\subsection{Work to failure increase} \label{sec:work_to_failure}

\begin{figure}[!h]
    \centering
    \includegraphics[width=1\linewidth]{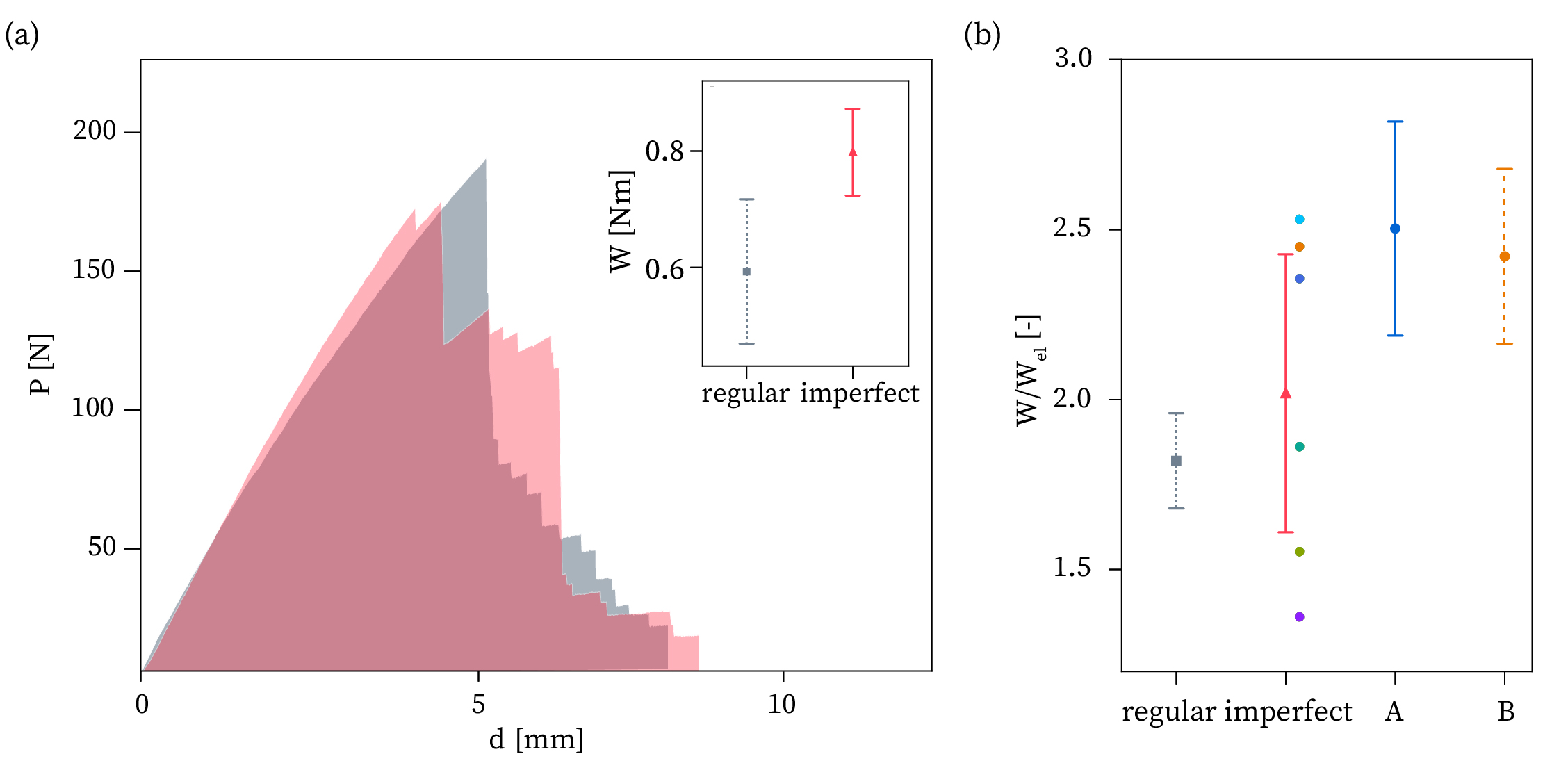}
    \caption{Comparison of work to failure of regular and imperfect specimens reveals toughening caused by imperfections. 
    (a)~Force-displacement curves of representative cases of regular and imperfect specimens, with the work to failure $W$ highlighted by the colored area. (inset) Statistical comparison of work to failure of all tested regular and imperfect specimens with 90\% confidence intervals. (b)~Work to failure normalized by the elastic contribution to work $W_{el}$ for the regular and imperfect lattices. All individual values are shown for the imperfect cases. Repetitions for configurations A (blue) and B (orange) with four and six specimens, respectively, are shown separately. All error bars correspond to 90\% confidence intervals.
    }
    \label{fig:3}
\end{figure}

As previously discussed, neither the regular nor the imperfect specimens fail completely right after the peak load. To quantify the entire energy absorption capability of the lattice, we compute the work to failure, $W$, which corresponds to the integral of the force-displacement curve
\begin{equation}
    W = \int_0^{d_\mathrm{u}} P(d) ~\mathrm{d}d ~,
    \label{eq:work_to_failure}
\end{equation}
where $P(d)$ is the measured force at a given displacement, $d$, and $d_\mathrm{u}$ is the ultimate displacement at failure~(Fig.~\ref{fig:2}a). The average work to failure of imperfect specimens systematically exceeds that of regular specimens (see Fig.~\ref{fig:3}a-inset). However, the high stress-intensity-factor fluctuations highlighted in Fig.~\ref{fig:2}c suggest that the peak load variability may bias the comparison of $W$. 

To ensure a fair assessment of the effects of imperfections on the work required to propagate the crack within regular and imperfect lattices, it is thus essential to normalize $W$ by its elastic contribution ($W_{el}$) up to the peak load. 
While the average normalized work to failure of imperfect specimens remains higher compared to the regular one, it is no longer a statistically relevant increase (see confidence intervals in Fig.~\ref{fig:3}b). By considering the individual values of $W/W_{el}$ for all tested imperfect lattice configurations, it is observed that three imperfect configurations outperform the regular lattice in terms of work to failure. 

To evaluate the consistency of these results, we select two of these three configurations, namely A and B, and repeat manufacturing and testing to verify the statistical robustness of their high work to failure. These additional results reveal that the normalized work to failure of these configurations is systematically increased compared to the regular case, as shown by the 90\% confidence interval in Fig.~\ref{fig:3}b. The average normalized work to failure for configuration A and B specimens is approximately 50\% and 40\% higher than that of the regular lattices, respectively. The consistency and significance of the increase in work to failure for configuration A and B specimens provides a strong basis for further analysis of the underlying mechanisms responsible for this observation, which will be performed in the following section.

\section{Local crack propagation} \label{sec:crack_path_characteristics}

We examine the observed increase in work-to-failure for configuration A and B specimens with respect to changes in the local failure mechanisms by analyzing the crack path. Identifying the crack path and tip from the acquired images (Fig.~\ref{fig:4}a) is challenging due to the high slenderness of the lattice beams and the resulting variations in gray levels. As an alternative, post-mortem inspection of the specimens (Fig.~\ref{fig:4}b) presents its issues, as additional damage may occur during specimen removal from the testing machine. To overcome these challenges, we employ a DIC-based approach, which systematically identifies the crack morphology and tip position directly from the images acquired during mechanical testing.

\subsection{DIC to unravel the crack path morphology} \label{sec:DIC_approach}

\begin{figure*}[!t]
    \centering
    \includegraphics[width=.95\linewidth]{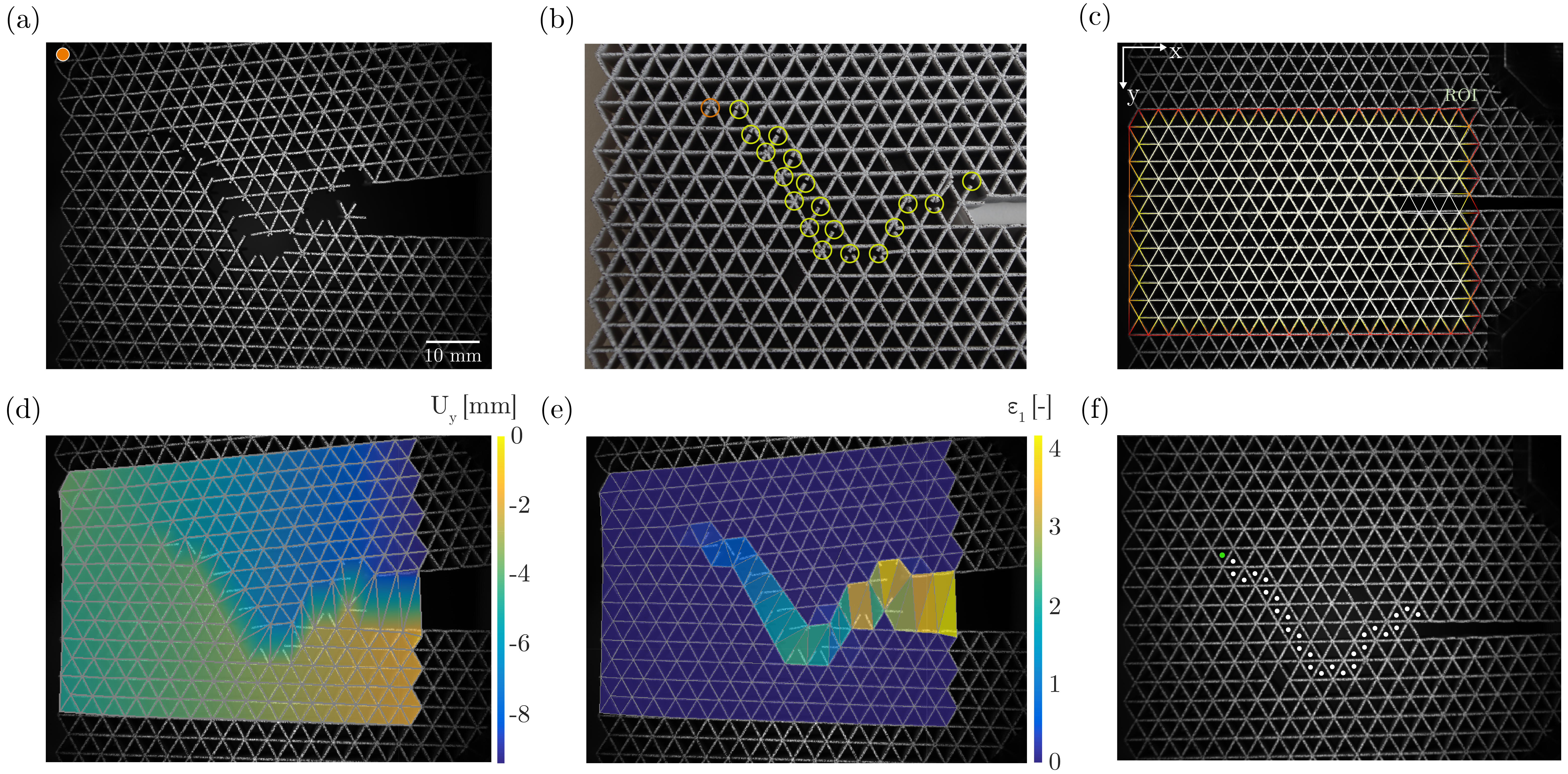}
    \caption{Automated detection of the crack path morphology by DIC for a representative imperfect specimen (configuration B). (a)~Image at the end of the mechanical test but before specimen removal illustrating a complex crack path morphology. (b)~Post-mortem image of the same specimen as in sub-figure (a). Yellow circles indicate detected lattice damage and the orange circle marks the deduced crack tip position. (c)~Region of interest (ROI) meshed with triangular elements. (d)~Vertical displacements field $U_y$ on the deformed mesh demonstrating an accurate analysis. (e)~Elementary maximum principal strain ($\varepsilon_1$) field drawn on the deformed mesh for the last acquired image shown in sub-figure (a). (f)~Identified crack path based on an elementary strain threshold. The centroid of the elements considered broken are marked with white dots and the crack tip is highlighted by a green marker.
    }
    \label{fig:4}
\end{figure*}

For DIC purposes, we pattern the specimen surface with Dupli-Color white and black spray paint. The high-resolution LIMESS camera position is adjusted to monitor a ROI that includes the notch root and the opposite free edge of the specimen for mesh backtracking purposes. The images acquired during the tests are processed using the finite-element-based DIC code Correli~3.0~\cite{correli} using a triangular mesh whose elements correspond exactly to the cells of the lattice specimen (see Fig.~\ref{fig:4}c). DIC exploits the contrast between the patterned specimen and the dark background, and minimizes the gray-level residual fields
\begin{equation}
    \rho_{\textbf{u}}(\textbf{x}) = f(\textbf{x}) - g(\textbf{x} + \textbf{U}(\textbf{x})) ~,
\end{equation}
where $f$ and $g$ are the reference and deformed images, respectively~\cite{Hild_2021}, and $\textbf{U}$ is the in-plane displacement field (Fig.~\ref{fig:4}d), which is measured through this approach. It is worth noting that the mesh is backtracked to ensure that the elements deform during loading and the node positions remain aligned with the lattice joints (see gray lines representing the elements in Fig.~\ref{fig:4}d). The backtracking procedure involves the generation of a mask, representing the nominal configuration, from the lattice STL file. This mask is then cropped to match the ROI. An initial correlation is performed between the image of the unloaded specimen and the mask to determine the nodal displacements. To backtrack the DIC mesh, the inverse of these nodal displacements is applied, effectively mapping the mesh from the nominal configuration to the reference configuration, which represents the undeformed state of the specimen.   

To determine the crack path and tip, the displacement fields obtained from DIC analyses are post-processed. We compute the maximum principal strain ($\varepsilon_1$) in each element, which reveals localized regions of higher levels along the crack path at the end of the test (see Fig.~\ref{fig:4}e). To identify broken lattice cells and locate the crack tip, we select a strain threshold of $\varepsilon_\mathrm{th} = 0.27$, which effectively distinguishes broken from intact cells. A qualitative comparison of the original image at the end of the test (Fig.~\ref{fig:4}a), the post-mortem analysis (Fig.~\ref{fig:4}b), and the DIC-based detection (Fig.~\ref{fig:4}f) across all specimens shows good agreement, thereby indicating that the DIC-based approach captures the crack path morphology and tip location reliably and with minimum effort.

\subsection{Characterization of the crack path} \label{sec:crack_path_characterization}

\begin{figure*}[!h]
    \centering
    \includegraphics[width=1\linewidth]{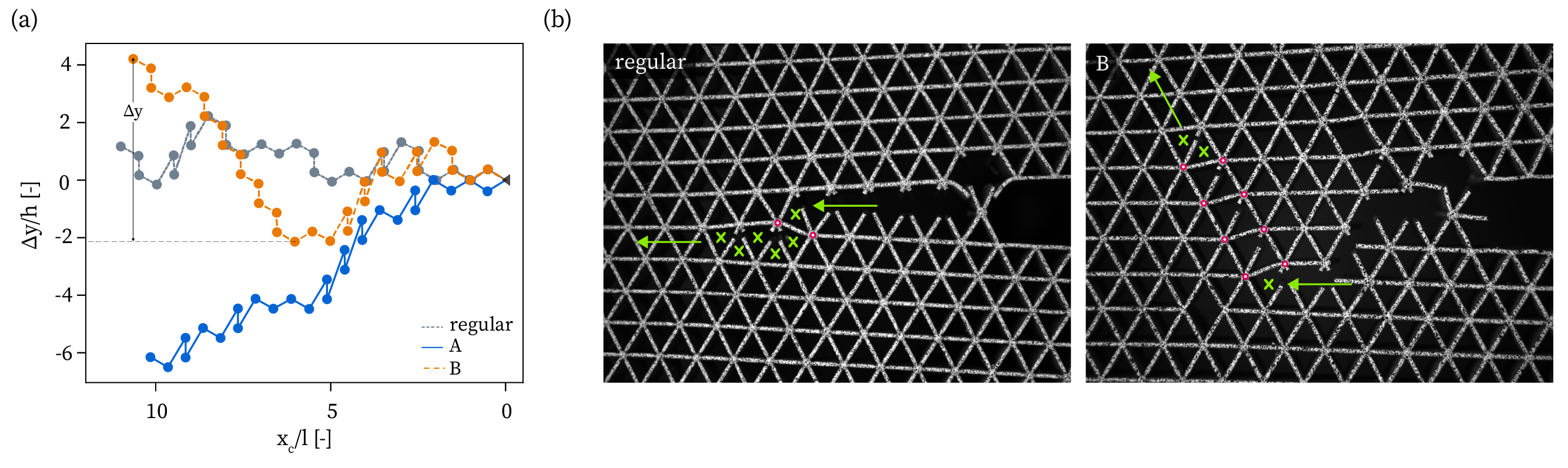}
    \caption{Characterization of the crack path by crack path morphology and crack bridging. (a)~Crack path derived from connecting the centroids of the elements whose strain exceeds a defined threshold. The $x$ and $y$ axes are normalized by the lattice cell width $l$ and height $h=l\sqrt{3}/2$, respectively. Lines connecting the centroids are included as visual guides only. A gray triangle at the origin marks the notch tip. A black arrow indicates the maximum crack path deviation $\Delta_y$ for configuration B. (b)~The direct inspection of the images reveals the presence of bridging struts across crack faces. The bridged crack path is highlighted with green crosses, and the nodes of the intact struts are marked in red.
    }
    \label{fig:5}
\end{figure*}

Considering three representative examples, one for each configuration, it is observed that the crack path in the regular specimen displays only minor deviations from a (relatively) straight trajectory (Fig.~\ref{fig:5}a). Conversely, the crack paths in the imperfect specimens show more pronounced deviations. In configuration A, the crack progressively drifts away from the original crack plane, without any major changes in direction. In configuration B specimen, the crack propagation direction changes twice, resulting in a more complex path (Fig.~\ref{fig:5}a). It is important to note that while all test specimens of any configuration are manufactured and tested in exactly the same manner, the observed crack path is not exactly identical, probably due to material or process-induced defects and environmental variabilities.

To systematically characterize the crack path morphology, a simple geometric approximation is introduced for the crack path tortuosity, namely, the maximum crack path deviation $\Delta_y$, defined as the difference between the maximum and minimum $y$-positions along the crack path (Fig.~\ref{fig:5}a). The average tortuosity normalized by the cell height for configuration A specimens (6.7~$\pm$~0.6) exceeds that of regular (2.8~$\pm$~1.2) and configuration B (3.5~$\pm$~1.3) specimens.
Furthermore, we observe that during fracture propagation, a few struts behind the crack tip remain intact across the crack faces, creating bridging (see Fig.~\ref{fig:5}b). We extract the number of bridging struts from the high-resolution images acquired during propagation and find that this number for configuration B (2.5~$\pm$~0.7) consistently exceeds that of regular (1.25~$\pm$~0.3) and configuration A (1.5~$\pm$~0.7) specimens. 

\section{Linking the crack path to the work to failure} \label{sec:linking_crack_path_to_wtf}

\begin{figure*}[!b]
    \centering
    \includegraphics[width=.8\linewidth]{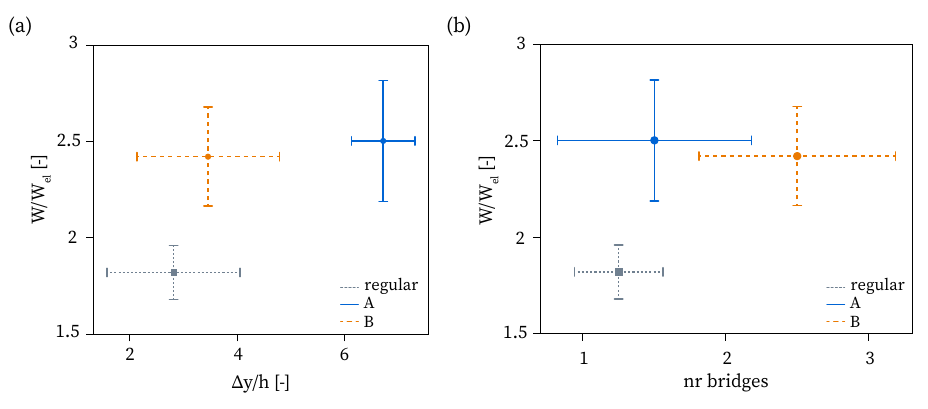}
    \caption{The work to failure ($W$) correlates with crack path tortuosity and bridging. $W$ as a function of (a)~maximum tortuosity (normalized by the cell height) and (b)~number of bridging struts, for regular and imperfect specimens. We normalize the work to failure with its elastic contribution ($W_{el}$).
    For all the measured quantities the error bars correspond to the 90\% confidence interval.
    }
    \label{fig:6}
\end{figure*}

By examining the correlation between the work to failure and the crack path tortuosity ($\Delta_y/h$), it is found that the higher crack path deviation observed for configuration A specimens consistently correlates with enhanced work to failure compared to regular triangular lattices (Fig.~\ref{fig:6}a). This observation supports the interpretation that the deviations in crack trajectory induced by the imperfections contribute to increased work to failure~\cite{Urabe, BOWER1991815}. Conversely, configuration B shows only a slight, statistically insignificant, increase in $\Delta_y/h$, thereby suggesting that other mechanisms may contribute to the increased work to failure for this configuration (Fig.~\ref{fig:6}b).

The work to failure gain for configuration B specimens correlates with the higher number of bridging struts during crack propagation compared to regular specimens (Fig.~\ref{fig:6}b). Conversely, the wide confidence interval characterizing configuration A specimens does not allow any effect of bridging to be distinguished for this configuration of imperfect lattices. These observations suggest that both crack path tortuosity and crack bridging act as toughening mechanisms in imperfect brittle lattices, with either mechanism prevailing depending on the specific location of the defect.

\section{Discussion} \label{sec:discussion}

A key observation of this study is that the presence of a few vacancies alters the crack propagation path, thereby increasing the work to failure in imperfect lattice specimens. In configuration A specimens, the removal of struts consistently led to a more tortuous crack path, as reflected by the increase in its maximum deviation from the notch. These deviations may be the reason for higher work to failure as they increase the effective fracture surface relative to the crack length, thereby requiring more energy to propagate the crack~\cite{Urabe, faber_evans_1983}. In configuration B specimens, the increase in work-to-failure correlates with the number of bridging struts. In analogy to the failure of composites~\cite{BOWER1991815}, the simultaneous presence of load-bearing struts behind the crack tip is expected to contribute to greater crack growth resistance. 
These findings suggest that crack path modifications, induced by local imperfections, serve as a mechanism to delay failure in brittle architected materials.
Another potential toughening mechanism that we did not consider in this study is crack pinning or temporary arrest at or near the location of the defect, which could delay propagation and result in a more extended loading phase~\cite{sanner2025,xiong_2024, kendall1975}. However, investigating this mechanism requires tracking the crack-tip advancement at each time step during testing, which is only possible by performing DIC analyses over a discrete lattice mesh and thus warrants further investigations. 

It is worth highlighting that, in our experiments, the presence of a small number of missing struts did not affect the location or mechanism of crack initiation. In all tested specimens, both regular and imperfect, fracture consistently initiated at the notch root, where the stress concentration is expected to be highest. This observation indicates that, for the specific six configurations considered herein, local imperfections placed one unit cell or more away from the notch root do not significantly interfere with the initiation process. This result is not necessarily general, as some imperfections located closer to the notch root could modify the local stress/strain fields and potentially influence the onset of failure.

A systematic sensitivity analysis examining how the increase in work to failure varies with the position of the removed strut, as well as with the number of removed struts, has not been performed owing to the complexity of the experimental procedures. A systematic variation of the placement of a single defect in spring network simulations~\cite{sanner2025} shows that a missing spring always increases or maintains the toughness (but never reduces it). However, the crack paths predicted by this spring network model do not agree with the paths observed experimentally, highlighting that more detailed simulation models are required to predict the failure of architected materials~\cite{fulco2025, Kraschewski_2024,dewaal2025}. 

%
%

%
%
%
%
%

%

%
%
\begin{comment}
This study focused on two specific imperfection configurations, both involving a small number of missing struts placed away from the notch root. While this choice allows for isolating and interpreting individual effects, the conclusions do not capture the full range of possible imperfection geometries or densities. In addition, the metric used to quantify crack path tortuosity is simple and does not account for the curvature or spatial complexity of the fracture trajectory. More refined geometric or energetic measures may reveal additional mechanisms or distinctions between configurations. Future work could explore a broader set of defect types, positions, and numbers to assess more systematically their influence on the fracture behavior. Combining such experimental studies with simulations could help disentangle local stress redistributions, crack pinning, and bridging effects. Extending the approach to different lattice topologies or loading modes would also clarify how general these observations are and inform design strategies for tougher, more failure-resistant architected materials.
\end{comment}

\section{Conclusion} \label{sec:conclusion}
In this work, we investigated the impact of missing struts on the failure of pre-notched triangular lattices. Our results indicate that a few imperfections enhance the work to failure. Using a DIC-based approach to track crack propagation, we associated this improvement with increased path tortuosity and bridging. These findings experimentally demonstrate that introducing disorder to enhance toughness is also effective for brittle lattices. 
Future studies should explore a wider range of defect types and locations, and integrate experiments with simulations to fully exploit defect engineering for fracture-resistant design.

\section{Acknowledgements}
The authors acknowledge support from the Swiss National Science Foundation under the SNSF starting Grant (TMSGI2\_211655). A.S. was supported by the Swiss National Science Foundation under grant number 200021\_200343. The authors thank Dr.~Mohit Pundir, Luca Michel, Flavio Lorez, Dr.~Jan Van Dokkum, Dr.~Leo De Waal, and Prof.~Marcelo Dias for fruitful discussions and feedback.

\section{CRediT authorship contribution statement}
\textbf{Alessandra Lingua}: Conceptualization, Methodology, Investigation, Formal Analysis, Data Curation, Visualization, Writing -- Original Draft. \textbf{Antoine Sanner}: Conceptualization, Writing -- Review \& Editing. \textbf{François Hild}: Methodology, Formal Analysis, Writing -- Review \& Editing. \textbf{David S. Kammer}: Conceptualization, Supervision, Writing -- Review \& Editing, Funding acquisition

\section{Declaration of competing interest}
The authors declare that they have no known competing financial interests or personal relationships that could have appeared to influence the work reported in this paper.

\section{Data Availability}
The data are available on the ETH research collection: [DOI]

\appendix
\nolinenumbers

\end{document}